\documentclass[12pt,paper]{JHEP3}
\usepackage{amsmath}
\usepackage{amssymb}
\usepackage{epsfig}
\newcommand{\be}{\begin{equation}}
\newcommand{\ee}{\end{equation}}
\newcommand{\bea}{\begin{eqnarray}}
\newcommand{\eea}{\end{eqnarray}}
\def\eqa{&=&}
\def\ccr{\nonumber\\}
\def\la{\langle}
\def\ra{\rangle}

\def\bchi{\bar{\chi}}
\def\bpsi{\bar{\psi}}
\def\btheta{\bar{\theta}}
\def\bep{\bar{\epsilon}}
\def\ep{\epsilon}

\title{Worldline approach to vector and antisymmetric tensor fields II}

\author{Fiorenzo Bastianelli\\
Dipartimento  di Fisica, Universit{\`a} di Bologna 
and  INFN, Sezione di Bologna, Via Irnerio 46, I-40126 Bologna, Italy \\ 
E-mail: \email{bastianelli@bo.infn.it}} 
\author{Paolo Benincasa\\
Department of Applied Mathematics, University of Western Ontario,
Middlesex College, London, ON, Canada N6A 5B7\\
E-mail: \email{pbeninca@uwo.ca}}
\author{ Simone Giombi\\
C.N. Yang Institute for Theoretical Physics,  
State University of New York at Stony Brook, 
Stony Brook, NY 11794-3840, USA\\
E-mail: \email{sgiombi@insti.physics.sunysb.edu}}

\abstract{
We extend the worldline description of vector and antisymmetric tensor fields 
coupled to gravity to the massive case. In particular, we derive a worldline 
path integral representation for the one-loop effective action of a massive 
antisymmetric tensor field of rank $p$ (a massive $p$-form) whose dynamics is 
dictated by a standard Proca-like lagrangian coupled to a background metric. 
This effective action can be computed in a proper time expansion to obtain the 
corresponding Seeley-DeWitt coefficients $a_0$, $a_1$, $a_2$. The 
worldline approach immediately shows that these coefficients are 
derived from the massless ones by the simple shift $D \to D+1$, where $D$ is 
the spacetime dimension. Also, the worldline representation makes it simple to 
derive exact duality relations. Finally, we use such a representation to 
calculate the one-loop contribution to the graviton self-energy due to both 
massless and massive antisymmetric tensor fields of arbitrary rank, 
generalizing results already known for the massless spin 1 field (the photon).}


\preprint{YITP-SB-05-24\\
          UWO-TH-05/07}

\begin{document}

\section{Introduction}
\label{sec:intro}

In a previous paper \cite{Bastianelli:2005vk} we have studied a worldline 
approach to massless antisymmetric tensor fields of arbitrary rank. 
In the present paper we extend such an approach to the massive case and use 
it: {\em (i)}
to identify the Seeley-DeWitt coefficients $a_0$, $a_1$ and $a_2$ for massive
antisymmetric tensor fields of arbitrary rank in  arbitrary dimensions, 
{\em (ii)} to study duality relations, and {\em (iii)} to compute
the one-loop contribution to the graviton self-energy 
(or equivalently, the stress tensor two-point function)
due to both massless and massive antisymmetric tensors.
This latter result generalizes the calculation 
for the massless spin 1 field (the photon)
\cite{Capper:1974ed} to the case of 
massless and massive antisymmetric tensor fields of arbitrary rank.

We denote our previous work \cite{Bastianelli:2005vk} as paper I, 
and refer to that paper for a general introduction to the subject.
Thus, in section 2 we directly discuss the $N=2$ spinning particle model
appropriate for describing massive antisymmetric tensor fields in first 
quantization.
In section 3 we construct the corresponding path integral on the 
circle (one-dimensional torus) to obtain the worldline representation 
of the effective action of a massive $p$-form. In this section we show that 
the Seeley-DeWitt coefficients for massive antisymmetric tensors
are related to the ones for  massless tensors by 
the simple shift $D \to D+1$, where $D$ is the spacetime dimension.
Thus we can readily compute them.
We also discuss duality relations between 
massive $p$- and $(D-p-1)$-forms and, 
at the level of unregulated effective actions, 
find that there is a mismatch 
proportional to the integral of the Euler topological density of 
target spacetime.
In section 4 we extract  the 2-point function from the effective action. 
It describes the 1PI one-loop graphs contributing to the graviton self-energy.
This generalizes  the worldline calculations for scalars 
and Dirac fermions ($N=0$ and $N=1$ spinning particles)
\cite{Bastianelli:2002fv,Bastianelli:2002qw}
to the case of massless and massive
antisymmetric tensors of arbitrary rank
($N=2$ spinning particle).
As a special case we recover the result for the massless photon
computed in \cite{Capper:1974ed}. 
In section 5 we point out a relation of the worldline 
representation of the effective action 
with the QFT description of massive 
tensors given in terms of a gauge invariant lagrangian with 
St\"uckelberg fields.
Then we briefly state our conclusions.
We use the appendix to explicitly list the 
Seeley-DeWitt coefficients $a_0$, $a_1$ and $a_2$ for 
massive $p$-forms with $p=0,1,2,3,4,5$
in arbitrary $D$ dimensions,  and to describe 
in some details the calculation of the two-point function.

\section{Massive $N=2$ spinning particle action}
The $N=2$ spinning particle can be used to describe massless and
massive antisymmetric tensors in first quantization
\cite{Brink:1976uf,Berezin:1976eg,Gershun:1979fb,Howe:1988ft,Howe:1989vn}.
In paper I  we reviewed the massless case. The inclusion of a mass term
is straightforward, and can be obtained by considering the massless case
in one dimension higher. In fact, one
can use the massless spinning particle in $D+1$ dimensions,
and then eliminate one dimension (for example $x^5$) by setting $p_5 =m$. 
The coordinate $x^5$ can be dropped from the action 
(it appears only as a total derivative), while the corresponding $N=2$
fermionic partners are retained and denoted by $\theta$ and $\bar \theta$.
This procedure gives the massive $N=2$ action, which in
phase space reads
\bea
S \eqa \int dt 
\Big [ p_\mu \dot x^\mu + i  \bpsi_\mu \dot \psi^\mu +
i \btheta \dot \theta 
- e H - i \bchi Q- i \chi \bar Q -  a J  + q a \Big ]  
\eea
where the constraints $C \equiv (H,Q,\bar Q, J)$  are given by
\bea
H = {1\over 2} (p_\mu p^\mu  + m^2)\ , \ \
Q = p_\mu \psi^\mu + m \theta \ , \ \
\bar Q = p_\mu \bar \psi^\mu + m \btheta \ , \ \
J =  \bar \psi^\mu \psi_\mu  +\btheta\theta \ . \quad\
\eea
They satisfy the Poisson brackets algebra (the $N=2$ algebra)
\bea
\{Q,\bar Q\}_{PB} = -2i H \ , \quad
\{J,Q\}_{PB} = i Q  \ , \quad
\{J,\bar Q\}_{PB} = - i \bar Q  
\label{algebra}
\eea
and are gauged by the gauge fields $G \equiv (e, \bar \chi, \chi, a)$. 
The quantized Chern-Simons coupling $q \equiv {D+1\over 2}-p-1$
allows to describe an antisymmetric tensor of rank $p$ 
\cite{Howe:1989vn}.
This is immediately seen in canonical quantization.
The phase space variables are turned into operators
satisfying the (anti)commutation relations (we use $\hbar=1$)
\be
[ \hat x^\mu , \hat p_\nu ] = i \delta^\mu_\nu \ ,
\quad \quad
\{ \hat \psi^\mu, \hat \psi^\dagger_\nu \} = \delta^\mu_\nu 
\ ,
\quad \quad
\{ \hat \theta, \hat \theta^\dagger \} = 1  \ .
\ee
States of the Hilbert space can be described by functions
of the coordinates $x^\mu$, $\psi^\mu$, $\theta$
(more precisely, one uses coherent states for the fermions),
and since $\psi^\mu$ and $\theta$ are Grassmann variables, 
the wave function has the following general expansion
\bea
\phi(x,\psi,\theta)
\eqa  F(x) + F_\mu(x) \psi^\mu + {1\over 2}
F_{\mu_1\mu_2}(x) \psi^{\mu_1}\psi^{\mu_2} +\ldots +
{1\over D!} F_{\mu_1\ldots\mu_D}(x) \psi^{\mu_1}\cdots \psi^{\mu_D} 
\ccr
&+& 
i m \Big ( A(x) \theta + A_\mu(x) \theta \psi^\mu +
{1\over 2} A_{\mu_1\mu_2}(x) \theta \psi^{\mu_1}\psi^{\mu_2} 
 +\ldots 
\ccr
&+& 
{1\over D!} A_{\mu_1\ldots\mu_D}(x) \theta \psi^{\mu_1}\cdots \psi^{\mu_D}
\Big)
\ .
\label{expa}
\eea
The imaginary unit $i$ makes it possible to impose reality conditions 
on the fields and the factor $m$ is introduced 
for obtaining a standard normalization of the $A$ fields.

The classical constraints $C$ become operators $\hat C$  
which are used to select the physical states through the requirements
$\hat C | \phi_{phys}\ra=0$,  
valid for all constraints except $\hat J$ which
gives $\hat J | \phi_{phys}\ra =q | \phi_{phys}\ra$
because of the Chern-Simons coupling. 
In the above coordinate representation these constraints 
 take the form of differential operators
\bea 
&& \hat H = {1\over 2}( -\partial_\mu \partial^\mu +m^2) \ , \ \
\hat Q = -i \psi^\mu \partial_\mu + m\theta 
\ , \ \
\hat Q^\dagger = -i \partial_\mu {\partial \over\partial \psi_\mu} 
+ m {\partial \over\partial \theta} 
  \ccr
&&
\hat J = - {1\over 2}\Big [\psi^\mu, {\partial \over\partial \psi^\mu} \Big ] 
- {1\over 2}\Big [\theta, {\partial \over\partial \theta}
 \Big ] 
= - \psi^\mu {\partial \over\partial \psi^\mu}
- \theta {\partial \over\partial \theta}   +{ D+1\over 2}
\eea
where in $\hat J$
 we have antisymmetrized $\hat \psi^\mu$, $\hat \psi^\dagger_\mu$
and $\hat \theta$, $\hat \theta^\dagger $
to resolve an ordering ambiguity.
The $\hat J=q$ constraint selects states with only 
$p+1$ Grassmann variables, namely
\be
\phi_{phys} (x,\psi) =
{1\over (p+1)!} F_{\mu_1\ldots\mu_{p+1}}(x) \psi^{\mu_1}\cdots 
\psi^{\mu_{p+1}} +
{i m \over p!} A_{\mu_1\ldots\mu_p}(x) \theta \psi^{\mu_1}\cdots \psi^{\mu_p}
 \ .
\ee
The constraints $\hat Q| \phi_{phys}\ra = 0$ gives the Bianchi identities
for $F_{p+1}$ and solves them in terms of $A_p$
\be
\partial_{[\mu} F_{\mu_1\mu_2\ldots\mu_{p+1}]} =0 \ , \quad 
 F_{\mu_1\mu_2\ldots\mu_{p+1}} =\partial_{[\mu_1} A_{\mu_2\ldots\mu_{p+1}]} 
\ee
while the constraint $\hat Q^\dagger| \phi_{phys}\ra =0 $ produces the Proca
equations together with the familiar longitudinal constraint on $A_p$
\be
\partial^{\mu_1} F_{\mu_1\ldots\mu_{p+1}}
= m^2 A_{\mu_2\ldots\mu_{p+1}} \ , \quad  
\partial^{\mu_1} A_{\mu_1\ldots\mu_{p}}=0  \ .
\ee
The constraint $\hat H| \phi_{phys}\ra =0 $ is
automatically satisfied as a consequence of the algebra 
$ \{ \hat Q , \hat Q^\dagger \} = 2 \hat H $. 

As these constraints reproduce the Proca field equations (Proca derived
them for a massive photon, but the generalization to an arbitrary $p$-form
is immediate), one may conclude that the massive $N=2$ spinning particle 
describes a massive antisymmetric tensor field with the standard Proca action
\be
S^{QFT}_p[A_p] = \int d^Dx \Big [-{1\over 2 (p+1)!} F_{\mu_1\ldots\mu_{p+1}}^2
-{m^2\over 2 p!} A_{\mu_1\ldots\mu_{p}}^2 \Big ] \ .
\ee
We now couple the massive spinning particle model to the target space metric,
go to configuration space, perform  a Wick rotation to euclidean time 
($t\to -i \tau$, and $ a\to i a$ to keep the gauge group $U(1)$ compact)
to obtain the euclidean action $ S_E$ that should be quantized
for describing in first quantization a massive
 antisymmetric tensor field of rank 
$p$ coupled to gravity. This action reads
\bea
S_E[\tilde X, G;g_{\mu\nu}] 
\eqa \int_0^1 d\tau 
\Big [ {1\over 2} e^{-1} g_{\mu\nu}
(\dot x^\mu - \bchi\psi^\mu- \chi\bpsi^\mu)
(\dot x^\nu - \bchi\psi^\nu- \chi\bpsi^\nu) 
\ccr
&+&  
\bpsi_a (\dot \psi^a + i a \psi^a 
+ \dot x^\mu \omega_\mu{}^{ab} \psi_b
)
-{e\over 2} R_{abcd} \bar\psi^a \psi^b \bar \psi^c \psi^d 
- i q a  
\ccr
&+&  \bar \theta (\dot \theta +i a \theta)
+im (\bar \chi \theta + \chi \bar \theta) + {1\over 2}e  m^2
\Big ] 
\label{fca}
\eea
where $\tilde X \equiv (x^\mu,\psi^a, \bpsi^a,\theta,\btheta)$
and $G \equiv (e, \bar \chi, \chi, a)$. 
We  recall that the parameter $q$ is  quantized as 
\be 
q={D+1\over 2} -p-1
\ee
in order to describe a massive $p$-form.
The gauge symmetries of the gravity multiplet on the worldline are unchanged
(they do not depend on the target space geometry)
\bea
\delta e \eqa \dot \xi + 2  \bchi \ep+ 2  \chi \bep \ccr
\delta \chi \eqa \dot \ep  + i a \epsilon - i \alpha   \chi \ccr
\delta \bchi \eqa \dot {\bep} -  i a  \bep + i \alpha  \bchi \ccr
\delta a \eqa \dot \alpha
\eea
and need to be gauge fixed to perform quantum calculations.

\section{Quantization on a torus}
The quantization on the one-dimensional torus goes through as in paper I, 
to which we refer for details.
Choosing the gauge $\hat G=(\beta, 0,0,\phi)$ for 
the supergravity 
multiplet, and inserting the factor $-1/2$ for 
correct overall normalization gives the following worldline representation
for the one-loop effective action of a massive $p$-form
\bea
\Gamma^{QFT}_{p}[g_{\mu\nu}] \eqa -{1\over 2}
\int_0^\infty {d \beta \over \beta}  \int_0^{2 \pi} {d \phi \over 2\pi}\,
\Big (2 \cos{\phi\over 2}\Big )^{-2} 
\int_{T^1} {\cal D} \tilde X\,  e^{-S_E[\tilde X, \hat G;g_{\mu\nu}]}
\label{34}
\eea
where $T^1$ denotes the one-dimensional torus (a circle),
and fermions have antiperiodic boundary conditions around the torus.

To present this effective action in a more explicit form, we integrate over 
$\theta$ and $\bar \theta$ which are free and give rise to an extra 
factor of the fermionic determinant $\det (\partial_\tau +i\phi) =
(2 \cos{\phi\over 2})$. Then we 
extract explicitly the normalization of the path integral measure
together with the bosonic zero modes $x^\mu_0$ to obtain 
\bea
\Gamma^{QFT}_{p}[g_{\mu\nu}] \eqa - {1\over 2}
\int_0^\infty {d \beta \over \beta} e^{-{1\over 2}m^2\beta}
\ccr
&& \times
\int_0^{2 \pi} {d \phi \over 2\pi}\,
\Big (2 \cos{\phi\over 2}\Big)^{D-1} e^{iq\phi}  
\ccr
&& \times
\int {d^D x_0 \sqrt{g (x_0)} \over (2 \pi \beta)^{D\over 2}}\, 
\big \la e^{-S_{int}} \big \ra  \ .
\label{act1}
\eea
The expectation values are normalized to one
(i.e. $\la 1 \ra =1$, with propagators given in paper I), and
the interactions are obtained from the nonlinear $N=2$ sigma model
\bea
S[X, \hat G;g_{\mu\nu}]
\eqa {1\over \beta} \int_0^1 d\tau \, 
\Big [ {1\over 2} g_{\mu\nu}(x) \dot x^\mu \dot x^\nu 
+   \bpsi_a (\dot \psi^a +  i \phi \psi^a +
\dot x^\mu \omega_\mu{}^a{}_b \psi^b) \ccr
&-& {1\over 2} R_{abcd} \bar\psi^a \psi^b \bar \psi^c \psi^d \Big ] 
\label{n2action}
\eea
which is the same as the one needed in the massless case
(by $X$ we denote the fields $(x^\mu, \psi^a, \bpsi^a)$).
The only difference from the massless case is the presence of the 
mass term $e^{-{1\over 2}m^2\beta}$  in the first line of 
(\ref{act1})
and the replacement $D\to D+1 $ which appears in the second line of 
(\ref{act1}).
This shows that it is quite simple to obtain the Seeley-DeWitt
coefficients in the massive case using our previous results 
for the massless case.
It is enough to shift $D\to D+1 $
in the result for the ``massless'' coefficients given in paper I
to obtain the  ``massive'' ones.
We list for convenience the explicit coefficients 
for the case $p=0,1,2,3,4,5$ in  appendix A1.
The general case is obtained by using the general formulas of paper I 
and shifting $D\to D+1 $.

Let us now discuss duality relations. 
To this end it is useful to recast (\ref{act1}) using 
the Wilson loop variable $w=e^{i\phi}$ and using an operatorial language 
\bea
\Gamma^{QFT}_{p}[g_{\mu\nu}] \eqa - {1\over 2}
\int_0^\infty {d \beta \over \beta}  e^{-{1\over 2}m^2\beta}
\oint_{\gamma} \frac{dw}{2 \pi i w} \,
\frac{w}{(1+w)^2}
{\rm Tr}\, \big [ w^{\hat N - p-1}
e^{-\beta  \hat H} \big ] \ccr
&\equiv&\!\!  -{1\over 2}
\int_0^\infty {d \beta \over \beta} e^{-{1\over 2}m^2\beta}
Z_p(\beta) \ .
\label{act2}
\eea
Here the path integral on the torus has been represented by the 
trace in the matter sector ($\tilde X$) 
of the Hilbert space of the spinning particle,
 $\hat N = \hat \psi^\mu \hat \psi^\dagger_\mu + \hat \theta 
\hat \theta^\dagger $ is the (anti) fermion number operator, 
and $\hat H$ is the quantum hamiltonian
without the coupling to the gauge field, which has been explicitly factorized.
As discussed in paper I, the regulated contour of integration for 
the $U(1)$ modular parameter $w$ is the unit circle  $\gamma$ 
with the point $w=-1$ excluded, see figure 1.
\[
\raisebox{-1cm}{\scalebox{.5}{
{\includegraphics*[1pt,1pt][290pt,290pt]{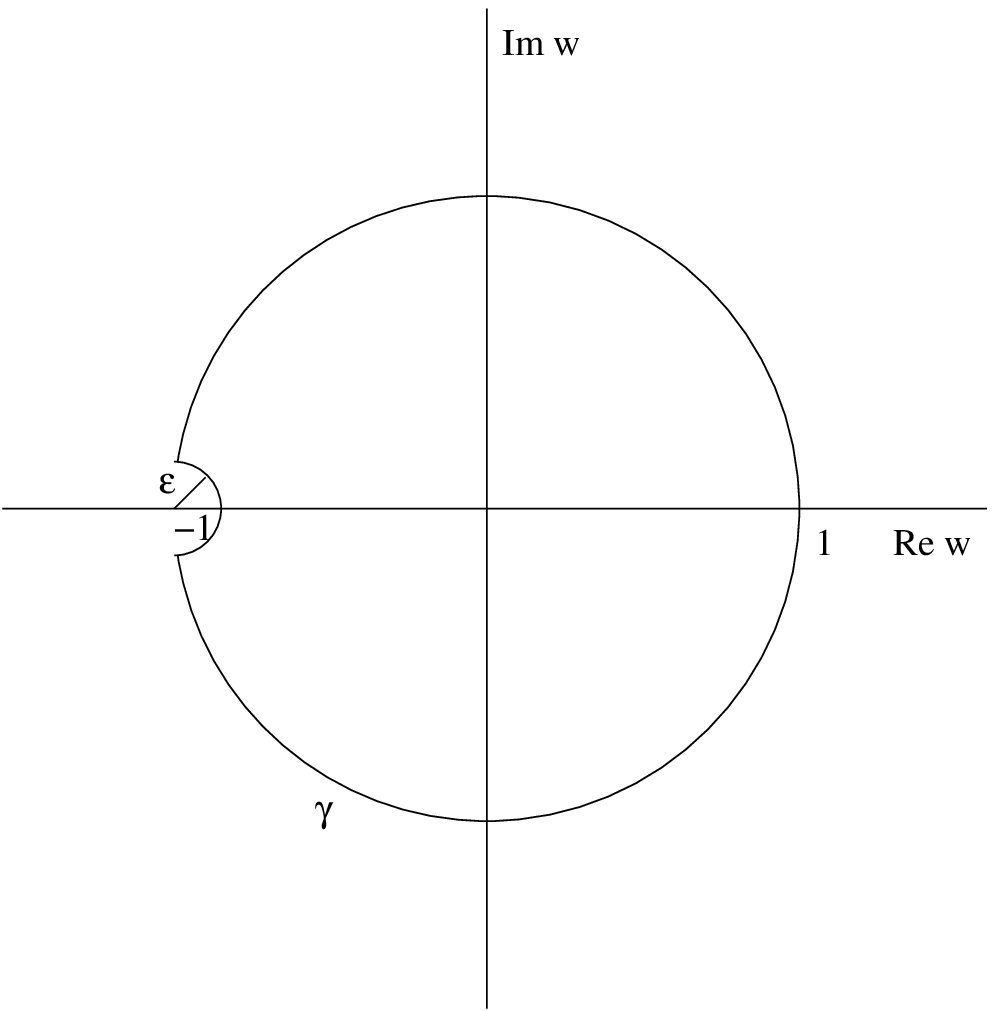}}}}
\]
\begin{center}
{\bf Figure 1:} Regulated contour in the variable $w$
\end{center}

One expects a massive $p$-form to be dual to a massive $(D-p-1)$-form.
Indeed changing $p\to (D-p-1)$ in (\ref{act1}) only changes the Chern-Simons 
coupling $q \to -q$. This can be undone by a
change of integration variable 
$\phi \to \phi' =-\phi$, or equivalently 
$w \to w'= {1\over w}$ in (\ref{act2}),
to go back to the original expression,
except that now the new contour $\gamma'$ is changed to
include the pole $w'=-1$, see figure 2.
\[
\raisebox{-1cm}{\scalebox{.5}{
{\includegraphics*[1pt,1pt][290pt,290pt]{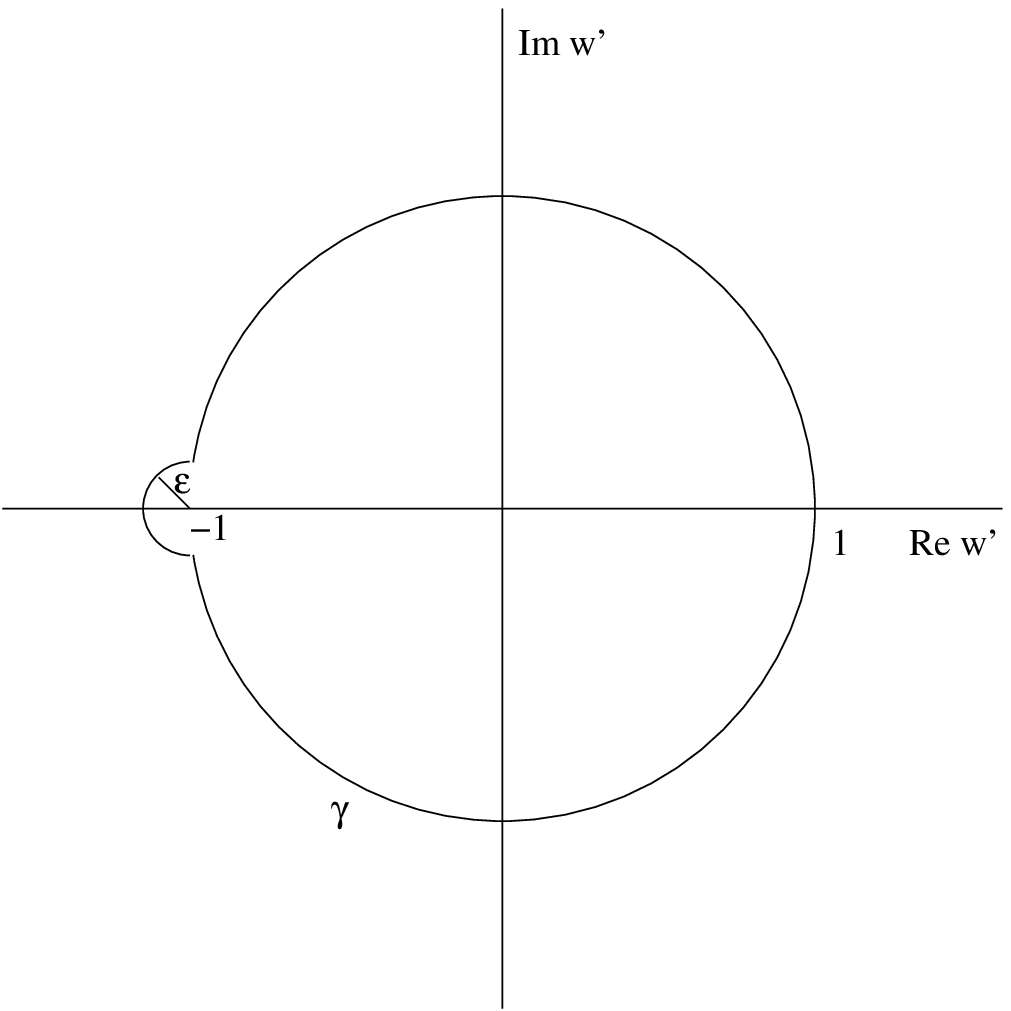}}}}
\]
\begin{center}
{\bf Figure 2:} Regulated contour in the variable $w'$
\end{center}
 
The mismatch $\Delta$ is easily computed as the residue at this pole
\bea
&&\Delta= 
{\rm Res}\left ( \frac{1}{(1+w)^2} 
{\rm Tr}\, [ w^{\hat N -p-1 }\,  e^{-\beta \hat H}], w=-1 
\right ) 
\ccr[1.8mm]
&& \quad \quad 
= {d\over d w} {\rm Tr}\, [ w^{\hat N -p -1 } e^{-\beta \hat H}]
\Big|_{w=-1} 
\ccr[1.8mm]
&& \quad \quad 
= {\rm Tr}\big [\big (\hat N -p-1\big ) 
\, (-1)^{\hat N -p- 2} \, e^{-\beta \hat H} \big ] 
\ccr[2mm]
&& \quad \quad 
=(-1)^{p}\, 
{\rm Tr}\, \big [ \hat N \, (-1)^{\hat N } \, e^{-\beta \hat H} \big ] 
+ (-1)^{p+1} (p+1)
{\rm Tr}\, \big [(-1)^{\hat N } \, e^{-\beta \hat H} \big ]
 \ccr[2mm]
&& \quad \quad = (-1)^{p+1}\, \chi
\eea 
where $\chi$ is the Euler character of target space.
In fact,
the trace ${\rm Tr}\, \big [(-1)^{\hat N } \, e^{-\beta \hat H} \big ] $
is seen to vanish, as the contributions from the
$F$ states are canceled by the contribution from the $A$
states (recall the expansion (\ref{expa})).
Similarly, one can compute
\be
{\rm Tr}\, \big [ \hat N \, (-1)^{\hat N } \, e^{-\beta \hat H} \big ] 
=-\chi \ .
\ee 
Thus one obtains the following exact duality relation
\be
Z_{D-p-1}(\beta) = Z_{p}(\beta) + (-1)^{p+1} \chi  
\ee
where $Z_{p}(\beta)$ is defined in (\ref{act2}).

Let us compare this result with the one for massless tensors.
The unregulated effective actions have a topological mismatch between
a massless $p$-form and the dual massless $(D-p-2)$-form. It is related
in even dimensions
to the topological Euler character \cite{Duff:1980qv,Bastianelli:2005vk},
and in odd dimensions to the Ray-Singer torsion 
\cite{Schwarz:1984wk}.
As we have just seen, a similar mismatch appears in even dimensions 
for a massive $p$-form and its dual massive $(D-p-1)$-form, and 
it is again proportional to the topological Euler character. 
It is worth noting that 
these mismatches are topological in nature, and are given by integrals 
of local expressions of the metric. 
They can be subtracted when the effective actions are renormalized, 
so that the renormalized effective actions can be defined 
to preserve duality, as pointed out in \cite{Siegel:1980ax}.
We will explicitly see this at the level of regulated
2-point functions discussed in the next section.

\section{Two-point function}

We now wish to use the $N=2$ spinning particle to compute the contribution 
to the graviton self-energy of massive and massless antisymmetric tensors of 
arbitrary rank in arbitrary spacetime dimensions. We choose to work in the 
vielbein basis, i.e. we take the 1-loop effective action to be a functional 
of the vielbein $\bar{\Gamma}^{QFT}_{p}[e_{a\rho}]
\equiv \Gamma^{QFT}_{p}[g_{\mu\nu}(e_{a\rho})]$. If desired, one can convert 
the final result to the metric basis using 
the appropriate change of variables (see eq.~(\ref{base})). 
Alternatively, one can do the computation 
directly in the metric basis by rewriting the $N=2$ spinning particle action 
in terms of worldline fermions with curved indices. In this setup, one should
also remember to add measure ghosts for the fermions as well 
\cite{Bastianelli:2002qw}. 

\[
\includegraphics*[100pt,667pt][241pt,725pt]{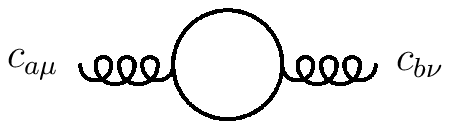}
\raisebox{-.5cm}{
\includegraphics*[100pt,652pt][221pt,725pt]{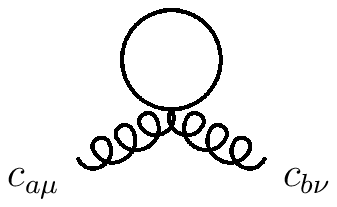}
}
\]
\vspace{-2mm}
\begin{center}
{\bf Figure 3:} One-loop contributions to the graviton self-energy. 
External lines represents vielbein fluctuations. 
\end{center}

The two-point function in configuration space can be obtained from the 
1-loop effective action by differentiating twice with respect to the vielbein 
(or metric) and then restricting to flat space. Alternatively, one can obtain
 the self-energy directly in momentum space by plugging into the worldline 
representation of the effective action, eq.~(\ref{34}), the following 
decomposition of the vielbein
\bea
e_{a\mu} &=& \delta_{a\mu} + c_{a\mu} \, \rightarrow \,  
g_{\mu \nu} = \delta_{\mu\nu} + 2 c_{(\mu\nu)} +c_{a\mu}c^a_\nu \nonumber \\
c_{a\mu} &=& \epsilon_{a\mu}^{(1)} e^{i k \cdot x(\tau)} + 
\epsilon_{a\mu}^{(2)} e^{-i k \cdot x(\tau)}
\label{vielbein}
\eea
and keeping only bilinear terms in the polarization tensors in the resulting 
calculation. In this way, one obtains the following expression for the 
two-point function
\bea
\bar{\Gamma}^{\mbox{\tiny $p$-form}}_{(k,-k)}[e_{a \nu}] &=& - 
{1\over 2 (2 \pi)^{D/2}}
\int_0^\infty {d \beta \over \beta^{1+D/2}} e^{-{1\over 2}m^2\beta} 
\int_0^{2 \pi} {d \phi \over 2\pi}\, \mu_{(m)}(\phi)
\int d^D x_0  \, 
\big \la e^{-S_{int}} \big \ra \Big|_{\mbox{\tiny m.l.}} \ 
\label{master}
\eea 
where $\mu_{(m)}(\phi)$ is the appropriate measure for the $U(1)$ modulus 
integration, i.e.
\bea
\mu_{(m)}(\phi) = \left \{ 
\begin{array}{ll}
\Big (2 \cos{\phi\over 2}\Big)^{D-2} e^{i(\frac{D}{2} -p-1 )\phi} 
\qquad & m = 0  \\ 
\\
\Big (2 \cos{\phi\over 2}\Big)^{D-1} e^{i (\frac{D+1}{2} -p-1 )\phi} 
\qquad & m \ne 0\  . \\
\end{array}
\right .
\label{measure}
\eea
The correlation function 
$\big \la e^{-S_{int}} \big \ra |_{\mbox{\tiny m.l.}}$ 
(the suffix here indicates that we 
keep only terms which are multi-linear in the polarization tensors) is the 
same for massive and massless tensors and is computed from the non-linear 
sigma model action in eq.~(\ref{n2action}). Of course, the appropriate measure 
ghosts must also be included \cite{Bastianelli:1991be,Bastianelli:1992ct}.

To carry out the explicit calculation, we find it convenient to use the 
string-inspired propagators. The coordinates
 $x$ are splitted as $x(\tau)=x_0 + y(\tau)$, where $y(\tau)$ satisfies 
periodic boundary conditions. 
The integration over the zero-mode $x_0$ gives momentum conservation, 
$(2\pi)^D \delta^D(\sum_i k_i)$, which we
 will not explicitly write in the following (and which we already used 
implicitly in eq.~(\ref{vielbein})). The detailed calculation of the 
correlation function in (\ref{master}) is reported in the appendix. Here we 
quote the final result, which can be written as 
\bea
\la e^{-S_{int}} \big \ra \Big|_{\mbox{\tiny m.l.}} &=& 
R_1 -{1 \over 2} (R_2-\bar{R}_2) + (A_0-1)(S_1+S_2) 
+{\beta k^2 \over 4} (2 A_0 -1) S_1 + 
{\beta^2 k^4 \over 16} S_1 A_0 \nonumber \\
&& + {1 \over 8 \mbox{cos}^2 {\phi \over 2} } 
(S_2-2 S_1) \left ( \beta k^2 (2 A_0-1) + {\beta^2 k^4 \over 4} A_0 \right ),
\label{final}
\eea 
where we defined the following set of tensors
\bea
R_1^{\mu\nu\alpha\beta} \!\! &=& \!
\delta^{\mu\nu}\delta^{\alpha\beta}
\nonumber \\
R_2^{\mu\nu\alpha\beta} \!\! &=& \!
\delta^{\mu\alpha}\delta^{\nu\beta}+ \delta^{\mu\beta}\delta^{\nu\alpha} 
\nonumber \\
\bar R_2^{\mu\nu\alpha\beta} \!\! &=& \!
\delta^{\mu\alpha}\delta^{\nu\beta}- \delta^{\mu\beta}\delta^{\nu\alpha}\ ,
\label{r-tensors}
\eea
along with the manifestly transverse tensors 
\bea
S_1^{\mu\nu\alpha\beta} \!\! &=& \!\!
\left(\delta^{\mu\nu}-{k^\mu k^\nu\over k^2}\right)
\left(\delta^{\alpha\beta}-{k^\alpha k^\beta\over k^2}\right) \nonumber \\
S_2^{\mu\nu\alpha\beta} \!\! &=& \!\!
\left(
\delta^{{\mu}{\alpha}}-{k^{{\mu}}
k^{{\alpha}}\over k^2}\right)\left(\delta^{{\nu}
{\beta}}-{k^{{\nu}}k^{{\beta}}\over k^2}\right) + (\mu \leftrightarrow \nu)
\ .
\label{s-tensors}
\eea
In eq.~(\ref{final}) we have suppressed the polarization tensors by denoting 
$R_i = \epsilon^{(1)}_{\mu\nu} \epsilon^{(2)}_{\alpha\beta} 
R_i^{\mu\nu\alpha\beta}$ and
$S_i = \epsilon^{(1)}_{\mu\nu} \epsilon^{(2)}_{\alpha\beta} 
S_i^{\mu\nu\alpha\beta}$, and we have also defined
\bea
A_0 = \int_0^1 d\tau e^{-{\beta \over 2}k^2 (\tau-\tau^2)}\ . 
\eea
The first line of eq.~(\ref{final}) is just the contribution of a 
minimally coupled scalar particle running in the loop, as can be 
checked by comparing with the results of \cite{Bastianelli:2002fv}
\footnote{In that paper the computation was performed in the metric basis.
 One can however use the relation between correlation functions in vielbein 
and metric basis, see eq.~(\ref{base}).}, and the second line can be viewed 
as the correction due to ``spin''.  

Inserting eq.~(\ref{final}) in eq.~(\ref{master}) and integrating over the 
moduli $\beta$ and $\phi$ we get the following expression
\bea
\bar{\Gamma}^{\mbox{\tiny $p$-form}}_{(k,-k)} &=& N_{dof} 
\bar{\Gamma}^{\mbox{\tiny scalar}}_{(k,-k)}
-{L_{(m)}(D,p) \over 8 (4\pi)^{D \over 2}} (S_2-2S_1)  \nonumber \\ 
&& \times \left ( k^2 \Gamma(1-{D \over 2})\big{(}2 (K^2)^{{D \over 2} -1} 
- (m^2)^{{D \over 2} -1}\big{)} 
+ {k^4 \over 2} \Gamma(2-{D \over 2}) (K^2)^{{D \over 2}-2} \right )\quad 
\label{2point}
\eea
where we have defined
\bea
(K^2)^x = \int_0^1 d\tau \left ( m^2+k^2 \tau (1-\tau)\right )^x .
\eea
The number of propagating degrees of freedom $N_{dof}$ is given by
\bea
N_{dof} = \int_0^{2 \pi} {d \phi \over 2\pi}\, \mu_{(m)}(\phi) = \left \{
\begin{array}{ccc}
{(D-2)! \over p! (D-2-p)!} \qquad m=0  \\
\\
{(D-1)! \over p! (D-1-p)!} \qquad m \ne 0 \\
\end{array}
\right .
\eea
and the coefficient $L_{(m)}(D,p)$ is equal to
\bea
L_{(m)}(D,p) =  \int_0^{2 \pi} {d \phi \over 2\pi}\, 
{ \mu_{(m)}(\phi) \over \mbox{cos}^2 {\phi \over 2} } 
\eea
or, in terms of the integrals $I_n(D,p)$ defined in paper I,
\bea
L_{(m)}(D,p) = \left \{
\begin{array}{ccc}
I_2(D,p)  \qquad m=0 \\
\\
I_2(D+1,p) \quad m \ne 0\ . \\
\end{array}
\right .
\eea 
All integrals over $\phi$ are regulated as prescribed 
in paper I (see also the previous section)
by going to the Wilson loop 
variable $w=e^{i \phi}$ and deforming the contour in the complex plane 
to exclude the pole at $w=-1$. Note that $L_{(m)}(D,0)=0$ for any $D$, so 
for $p=0$ the result reduces to the scalar contribution 
$\bar{\Gamma}^{\mbox{\tiny scalar}}_{(k,-k)}$ as it should. Explicitly, 
this contribution is obtained from first line of eq.~(\ref{final}) after 
integrating over the proper time $\beta$ and is given by
\bea
(4\pi)^{D \over 2}\bar{\Gamma}^{\mbox{\tiny scalar}}_{(k,-k)} \!&=&\! 
-{1 \over 2} \Gamma\Big (-{D \over 2}\Big ) \Bigg{(} (m^2)^{{D \over 2}} 
(R_1-{1 \over 2} (R_2-\bar{R}_2)) + 
((K^2)^{{D \over 2}}-(m^2)^{{D \over 2}})(S_1+S_2) \Bigg{)} \nonumber \\
&& \hskip -1cm
-{1 \over 4} \Gamma\Big (1-{D \over 2}\Big ) k^2 
\left ( 2(K^2)^{{D \over 2}-1}-(m^2)^{{D \over 2}-1} \right ) S_1
-{1 \over 8} \Gamma\Big (2-{D \over 2}\Big ) k^4  (K^2)^{{D \over 2}-2} S_1 \ .
\nonumber \\
\label{scalar}
\eea  

Eq.~(\ref{2point}) is our final result for the graviton self-energy in the
 vielbein basis. If one wishes to express the two-point function in the
 metric basis, one can use the following relation
\bea
\Gamma_{(k,-k)} [g_{\mu\nu}]= {1 \over 4} 
\left( \bar{\Gamma}_{(k,-k)}[e_{a\nu}] - \epsilon^{(1)}_{\mu\nu} 
\epsilon^{(2)}_{\alpha\beta}\delta^{\mu \alpha} 
\bar{\Gamma}^{\nu\beta}_{(0)}[e_{a\nu}]  \right )
\label{base}
\eea 
which depends also on the one-point function 
$\bar{\Gamma}^{\nu\beta}_{(0)}[e_{a\nu}]$. This relation 
simply follows from $\bar{\Gamma}^{QFT}_{p}[e_{a\rho}] = 
\Gamma^{QFT}_{p}[g_{\mu\nu}(e_{a\rho})]$ and $g_{\mu\nu} = 
\delta_{ab} e^a{}_\mu e^b{}_\nu$. In the present case the
one-point function is easily computed
\bea 
\epsilon^{(1)}_{\mu\nu} \epsilon^{(2)}_{\alpha\beta} 
\bar{\Gamma}^{\mbox{\tiny $p$-form}\, \nu\beta}_{(0)} \delta^{\mu \alpha} 
= -{N_{dof} \over 4 (4\pi)^{D \over 2}} (R_2 + 
\bar{R}_2) (m^2)^{{D \over 2}} \Gamma\Big (-{D \over 2}\Big )   \ .
\eea
It only affects the scalar contribution in eq.~(\ref{2point}), so the final 
expression for the two-point function in the metric basis is
\bea
\Gamma^{\mbox{\tiny $p$-form}}_{(k,-k)} &=& N_{dof}
 \Gamma^{\mbox{\tiny scalar}}_{(k,-k)}
-{L_{(m)}(D,p) \over 32 (4\pi)^{D \over 2}} (S_2-2S_1)  \nonumber \\ 
&& \times \left ( k^2 \Gamma\Big (1-{D \over 2}\Big )
\big{(}2 (K^2)^{{D \over 2} -1} 
- (m^2)^{{D \over 2} -1}\big{)} 
+ {k^4 \over 2} \Gamma\Big (2-{D \over 2}\Big ) 
(K^2)^{{D \over 2}-2} \right ) \quad\quad 
\label{2pointmet} 
\eea
where $\Gamma^{\mbox{\tiny scalar}}_{(k,-k)}$ is the two-point function 
due to a minimally coupled scalar in the metric 
basis, which was given in \cite{Bastianelli:2002fv} and is essentially the 
same as eq.~(\ref{scalar}) after dividing by 4 and replacing 
${1 \over 2} (R_2-\bar{R}_2) \rightarrow R_2$ (due to the tadpole correction 
in eq.~(\ref{base})).

As a consistency check, one can note that eq.~(\ref{2point}) and 
eq.~(\ref{2pointmet}) satisfy the gravitational 
Ward identities (which can be found in \cite{Bastianelli:2002fv}),
since the scalar
 contribution was already checked to satisfy them in 
\cite{Bastianelli:2002fv},
 and the additional correction is transversal and symmetric in each pair of
 indices (the one-point function contribution to the Ward identities only 
affects the scalar part).

We can now make several comments about the expected dualities of the final 
result for the two-point function. 
As noted in paper I for the massless case and in section 3 
for the massive case, the mismatch between the (unregulated) effective actions
 of dual forms is due to the appearance of a pole at $\phi=\pi$, 
or $w=-1$. It was also shown that the mismatch is always related to 
topological invariants of the target space, namely the Euler character
for even dimensions
(massless and massive case) and the Ray-Singer torsion for 
odd dimensions
(massless case only). Therefore one would expect that correlation 
functions, which are obtained differentiating the effective action with 
respect to the vielbein (or metric), do not show any mismatch even at the 
unrenormalized level. By looking at our final result, eq.~(\ref {final}) 
or eq.~(\ref{2pointmet}), and keeping in mind the measure factor for the 
integration over $\phi$ given in eq.~(\ref{measure}), we see that the only 
possible poles at $\phi=\pi$ occur for $D=2$ (both in massless and massive 
case) and $D=3$ for massless forms. These poles affect the coefficient 
$L_{(m)}(D,p)$ which will not respect duality in those special cases.
 However, in $D=2$ one has the identity $S_2=2S_1$, so the whole 
contribution proportional to $L_{(m)}(D,p)$ cancels out in this case. 
Moreover, one can verify that the expression in parenthesis in the second 
line of eq.~(\ref{2point}) turns out to be proportional to $(D-3)$ in the 
$m=0$ limit, and thus the term multiplying $L_{(m)}(D,p)$ cancels out also 
in the $D=3$ massless case. From these observations we may conclude that the 
two-point function, eq.~(\ref{2point}), 
respects  dualities in any dimension for 
both massive and massless forms. However, this is true formally at a given 
integer dimension $D$, but 
one has to be careful when applying dimensional regularization to regulate 
the proper time integral. In $D$ odd there are 
no poles in DR, and indeed the two-point functions respect duality without 
any mismatch as one can infer from the above observations. For even $D$ 
dimension, however, UV divergences arising from the proper time integral show 
up in DR as $1/\epsilon$ poles from the $\Gamma$-functions in 
eq.~(\ref{2point}). But one is allowed to conclude that the mismatch between 
dual forms is topological only at a fixed integer $D$, while in DR 
$\epsilon$-dependent terms which are not topological could appear. One 
can see such an effect by looking at the coefficients $N_{dof}$ and 
$L_{(m)}(D,p)$. In fact, although these respect duality at a given integer 
$D$, they loose this property in DR. For example, taking a massless scalar
 and a massless 2-form in four dimensions:
\bea
N_{dof} &=& 1 \qquad \qquad \qquad \qquad \quad p=0 \quad D=
4+\epsilon \nonumber \\ 
N_{dof} &=& 1 + {3 \over 2} \epsilon + {1 \over 2} \epsilon^2 
\qquad \qquad p=2 \quad D=4+\epsilon \ . 
\eea
Thus, because of  the $1 / \epsilon$ pole from the $\Gamma$-function, 
dual forms receives different finite contributions to the two-point functions.
 However, the divergent $1/ \epsilon$ part of the two-point function still 
respects strict duality as expected, since in this case one can keep 
everything at fixed integer $D$ where the mismatch is topological and is not 
expected to appear in correlation functions. This finite discrepancy between 
dual forms is due to DR: in other schemes, such as a
cutoff in proper time regularization, 
it should not appear. Of course, this finite mismatch will not affect the 
renormalized two-point function as it is local and 
one can always remove it in the renormalization procedure.     

\subsection{Massless $p$-form and the photon}
Setting $m=0$ in the final expression eq.~(\ref{2pointmet}), one gets 
the contribution to the graviton self-energy (in the metric basis) due to 
a massless $p$-form gauge potential 
\bea
\Gamma^{\mbox{\tiny $p$-form}}_{(k,-k)} \Big|_{m=0}  &=& N_{dof} 
\Gamma^{\mbox{\tiny scalar}}_{(k,-k)}\Big|_{m=0}  
- {D-3 \over D-1 }L_{(0)}(D,p){(k^2)^{{D \over 2}} \over 
64(4\pi)^{D \over 2}} 
\nonumber \\ 
&& \times
\Gamma \Big (2-{D \over 2} \Big) B\Big({D \over 2}-1,{D \over 2}-1\Big) 
 (S_2-2S_1)  \label{massless form}
\eea
where $B(x,y)=\Gamma(x)\Gamma(x)/\Gamma(x+y)$ is the Euler beta function.
The contribution of the massless scalar can be written as
\bea
\Gamma^{\mbox{\tiny scalar}}_{(k,-k)} \Big|_{m=0} &=& -{(k^2)^{{D \over 2}}
 \over 32(4\pi)^{D \over 2}} 
\Gamma\Big (2-{D \over 2}\Big ) B\Big ({D \over 2}-1,{D \over 2}-1\Big )
\nonumber \\ &&\times
 \bigg{[} {D^2-2D-2 \over D^2-1 }S_1 +  { 1 \over D^2-1 } S_2 \bigg{]} 
\ . 
\label{masslessscalar}
\eea
The graviton self-energy due to a photon loop was previously computed
 in \cite{Capper:1974ed} using standard QFT techniques. To compare with 
their result, one can set $p=1$ in the above general expression for
 massless tensors. In this case $N_{dof}=D-2$ and $L_{(0)}(D,1)=4$, 
so that we get
\bea
\Gamma^{\mbox{\tiny photon}}_{(k,-k)} &=&
 -{(k^2)^{{D \over 2}} \over 32(4\pi)^{D \over 2}}
\Gamma\Big (2-{D \over 2}\Big )B\Big ({D \over 2}-1,{D \over 2}-1\Big ) 
 \nonumber \\
&& \times \bigg{[} {D^3-8D^2+10D+16 \over D^2-1 } S_1 +
 {2D^2-3D-8 \over D^2-1} S_2 \bigg{]} \ .
\eea
One can check that this result agrees with the one 
in \cite{Capper:1974ed}. To make the comparison one has to keep in mind
 that in \cite{Capper:1974ed} correlation functions are defined by
 fluctuations of $\hat{g}^{\mu\nu} \equiv \sqrt{g}g^{\mu \nu}$, while
 we consider fluctuations of the metric $g_{\mu\nu}$.

\section{St\"uckelberg fields for massive tensors}

In this section we will show how the $N=2$ worldline description of massive 
tensors naturally suggests 
a way of rewriting of the QFT Proca action in terms of St\"uckelberg fields  
and an abelian gauge invariance with the relative ghosts for ghosts tower. 

We first review the case of massless tensors. As discussed in detail  
in paper~I, the massless $N=2$ spinning particle 
describes a  generalized field strength $F_{p+1}$ satisfying Maxwell 
equations and Bianchi  identities. 
This constraint is enforced by gauging the $N=2$ worldline supersymmetry. 
The net effect of this gauging 
is to produce a factor
 $(2 \mbox{cos} {\phi \over 2}) ^{-2}= {w \over (1+w)^2}$ 
coming from the susy ghosts which modifies the measure for the $U(1)$ modular 
integration. In fact, one can see that this factor coming from worldline 
susy ghosts automatically encodes the appropriate ghosts for ghosts tower 
needed to covariantly gauge fix
the gauge potential $A_p$. This is most easily seen by considering the 
operatorial representation of the partition function for the $p$-form gauge 
potential 
\bea
Z_p (\beta) = \oint_{\gamma} \frac{dw}{2 \pi i w} \,
\frac{w}{(1+w)^2}
{\rm Tr}\, \big [ w^{\hat N - p-1}
e^{-\beta  \hat H} \big ] = \oint_{\gamma} \frac{dw}{2 \pi i w} 
\frac{w}{(1+w)^2} \sum_{n=0}^D  w^{n- p -1}\, t_n(\beta)  
\nonumber \\
\label{Zmassless}
\eea
where $t_n(\beta)$ is ${\rm Tr}\, e^{-\beta \hat H}$ restricted to the space 
of $n$-forms (the sector of the Hilbert space with occupation number 
$\hat N=n$).
The pole at $w=-1$ is excluded, 
thus we only need to consider the pole at $w=0$. Around $w=0$, one 
has the expansion
\bea
{1 \over (1+w)^2} = 1-2w^2+3w^3-4w^4+ .\,.\,.
\eea
whose coefficients are seen to give the right multiplicity and statistics of 
the ghosts for ghosts tower. Indeed, plugging this expansion into 
eq.~(\ref{Zmassless}) and using Cauchy theorem one gets
\bea
Z_p (\beta) = \sum_{n=0}^p (-1)^n (n+1) t_{p-n} (\beta)
\eea
which corresponds to the correct triangular ghost structure 
for the $p$-form gauge potential.

We now turn to the massive case. As explained in section 2, gauging 
supersymmetry gives the Proca field equations for a massive tensor $A_p$.
 The worldline susy ghost factor now combines with the determinant arising
 by integrating out the 
fermionic $\theta$ variables, to give a net factor 
$(2 \mbox{cos} {\phi \over 2}) ^{-1}$ in the measure. Following the lines 
of the massless case, we expect that this factor could be viewed to create
 some kind of ``auxiliary field" tower for the 
massive Proca tensor. In fact, consider again the operatorial form of the 
partition function
\bea
Z_p(\beta) &=& \oint_{\gamma} \frac{dw}{2 \pi i w} \,
\frac{w}{(1+w)^2}
{\rm Tr}\, \big [ w^{\hat \psi^\mu \hat \psi^\dagger_\mu + \hat \theta 
\hat \theta^\dagger - p-1}
e^{-\beta  \hat H} \big ] \nonumber \\
&=& \oint_{\gamma} \frac{dw}{2 \pi i w} \,
\frac{w}{1+w}
{\rm Tr'}\, \big [ w^{\hat \psi^\mu \hat \psi^\dagger_\mu - p-1}
e^{-\beta  \hat H} \big ] \nonumber \\
&=& \oint_{\gamma} \frac{dw}{2 \pi i w} 
\frac{w}{1+w} \sum_{n=0}^D  w^{n- p -1}\, t_n(\beta)
\label{Zmassive}
\eea  
where first we have traced over eigenstates of $\hat \theta 
\hat \theta^\dagger$ to produce a factor $(1+w)$ (note that $\hat H$ does 
not depend on $\hat \theta$'s), and then expressed the 
remaining trace (Tr$'$) over 
eigenstates of $\hat \psi^\mu \hat \psi^\dagger_\mu$, i.e. the $n$-forms
of the restricted Hilbert space. 
Now the measure factor has the following expansion around $w=0$
\bea
{1 \over 1+w} = 1- w^2 + w^3 - w^4 + .\,.\,.
\eea
Plugging into eq.~(\ref{Zmassive}) we get
\bea
Z_p (\beta) &=& t_p(\beta) - t_{p-1} (\beta) + t_{p-2} (\beta) - .\,.\,. 
\nonumber \\ 
            &=& \sum_{n=0}^p (-1)^n t_{p-n} (\beta)\ .
\label{tower}
\eea
This suggests that the Proca action could be rewritten in terms of a tower 
of differential forms with simple kinetic operator (the 
generalized laplacian $d d^\dagger +d^\dagger d$).
 One can check that the above relation is correct by comparing the 
Seeley-DeWitt coefficients for massive forms given in appendix with the 
coefficients for differential forms with generalized 
laplacian as kinetic operator which
 were computed in paper~I. Also notice that 
eq.~(\ref{tower})
 is in agreement with the exact duality relations derived in section 3.

We now would like to interpret eq.~(\ref{tower}) from the QFT point of view.
 This will produce a useful way of rewriting the Proca action.
 For simplicity, consider $p=1$ (massive photon) and 
flat space. The generalization to curved space is straightforward. 
The Proca action is
\bea
S^{QFT} [A_{\mu}] = \int d^Dx \Big [-{1\over 4} F_{\mu\nu}^2
-{m^2\over 2 } A_{\mu}^2 \Big ] \ .
\eea
This action does not have any gauge invariance. However, we could 
add by hand a scalar field $\phi$ (a St\"uckelberg field) in 
the following way
\bea
S^{QFT} [A_{\mu},\phi] = \int d^Dx \Big [-{1\over 4} F_{\mu\nu}^2
-{m^2\over 2 } \Big 
(A_{\mu}-{1\over m} \partial_{\mu} \phi \Big )^2 \Big ] \ .
\eea  
Clearly now we have the gauge invariance
\bea
\delta A_{\mu} = \partial_{\mu} \, \Lambda \nonumber \\
\delta \phi = m \, \Lambda  \ .
\eea
Of course, $\phi$ could be completely gauged away from the action by using 
this symmetry. To our purpose it is convenient to 
keep it and choose a different gauge fixing. Following the standard BRST 
procedure, we introduce the ghost $c$, the non-minimal fields $(b,\pi)$,
 and the BRST transformations
\bea
\delta_B A_{\mu} &=& \partial_{\mu} \, c \nonumber \\
\delta_B \phi &=& m \, c \nonumber \\
\delta_B c &=& 0 \nonumber \\
\delta_B b &=& \pi \nonumber \\
\delta_B \pi &=& 0 \ .
\eea
Choosing the following gauge fermion
\bea
\Psi = b \, \Big (\partial_{\mu} A^{\mu} -m \, \phi  + \frac{1}{2} 
\pi \Big ) \, ,
\eea
the gauge fixed action gets the form
\bea
S^{QFT} [A_{\mu},\phi,b,c] &=& \int d^Dx \Big [-{1\over 4} F_{\mu\nu}^2
-{m^2\over 2 } (A_{\mu}-{1\over m} \partial_{\mu} \phi )^2  + 
\delta_B \Psi \Big ]  \\
&=& \int d^Dx \Big [ {1 \over 2} A_{\mu} (\partial^2 - m^2) A^{\mu} + 
{1 \over 2} \phi (\partial^2 - m^2) \phi 
- b (\partial^2 - m^2) c \Big ] \nonumber 
\eea
where in the last line the auxiliary field $\pi$ was integrated out. 
This shows that the massive photon is equivalent to 
the sum of a vector, a commuting scalar ($\phi$)
and two anticommuting scalars ($b,c$), 
all with the simplest kinetic operator. It is easy to see that the 
alternating tower of eq.~(\ref{tower}) is reproduced. 
One advantage of using these extra fields is to have simpler 
kinetic terms, and thus simpler propagators. A description of this 
St\"uckelberg field approach to  massive spin 1 fields can also be found in
the textbook \cite{Siegel:1999ew}.

It is straightforward to generalize this procedure for tensors of higher 
rank. For example, for the massive 2-form $A_2$ one adds a St\"uckelberg 
vector field $\phi_1$ in the Proca action through the substitution
\bea
A_2 \rightarrow A_2 - {1 \over m} d \phi_1 \ .
\eea
This introduces the gauge symmetries $\delta A_2 = d \Lambda_1$ and 
$\delta \phi_1 = m \Lambda_1 + d \Lambda_0 $. Upon gauge 
fixing one gets two anticommuting  vector ghosts and three commuting scalar 
ghosts from the $\Lambda_1$-symmetry, plus two anticommuting scalar ghosts 
from the $\Lambda_0$-symmetry. Therefore the alternating structure of 
eq.~(\ref{tower}) is again reproduced.

\section{Conclusions}

We have extended the worldline description of massless antisymmetric 
tensor fields coupled to gravity initiated in paper I 
to the  massive case, and used the worldline description to study 
various quantum effects. In particular, we have computed the one-loop 
effective action in its proper time expansion (a local derivative 
expansion) and obtained the first few  Seeley-DeWitt coefficients
($a_0$, $a_1$, $a_2$). This derivative expansion is 
really valid only for massive fields, because otherwise infrared divergences 
appear. However the Seeley-DeWitt coefficients are 
useful in the massless case as well: they identify, for example, 
the counterterms needed to renormalize the full effective action. 
Then we have studied exact duality relations between 
antisymmetric tensor fields and obtained the topological terms
that cause a mismatch in their unregulated effective action.
Finally we have computed the two-point function of the metric 
with an arbitrary rank $p$ antisymmetric tensor field in the loop, 
either massive or massless. For the particular case of a massless 1-form,
we have reproduced the result obtained in \cite{Capper:1974ed}.
The worldline perspective employed here and in paper I 
has been instrumental to obtain general results in the quantum theory 
of antisymmetric tensor fields, which may find some 
applications in theoretical particle and string physics.

\acknowledgments 

The research of FB has been sponsored in part by the Italian MIUR 
grant PRIN-2003023852 
``Physics of fundamental interactions: gauge theories, gravity and strings''
and by the EU network EUCLID (HPRN-CT-2002-00325).
The work of SG was partially supported by NSF grant no. PHY-0354776.
FB and PB are grateful to Stefan Theisen and the Albert Einstein Institute for 
the stimulating environment and hospitality. FB would like to thank 
Martin Ro{\v c}ek and the organizers of the Simons Workshop on Mathematics and 
Physics 2004 at SUNY at Stony Brook for a similar stimulating environment and 
hospitality.

\appendix
\section{Appendix}
\label{section:appendix}

\subsection{List of Seeley-DeWitt coefficients for massive $p$-forms $A_p$}

One may write the one-loop effective action for a massive $p$-form $A_p$
as follows
\bea
\Gamma^{QFT}_{p}[g_{\mu\nu}] 
\eqa -{1\over 2}
\int_0^\infty {d \beta \over \beta} e^{-{1\over 2}m^2\beta}\, Z_p(\beta) \ccr
Z_p(\beta) \eqa
\int {d^D x \sqrt{g} \over (2 \pi \beta)^{D\over 2}}\, 
\Big ( a_0 + a_1 \beta +a_2\beta^2 + \cdots \Big )
\eea
where the coefficients $a_i$ are the so-called Seeley-DeWitt coefficients
in the coincidence limit. We parametrize them as follows   
\be
\Big ( v_1 + v_2 R \beta 
+ (v_3 R^2_{abcd}+v_4 R_{ab}^2+ v_5 R^2 + 
v_6 \nabla^2 R)\beta^2  + \cdots 
\Big )  \ .
\ee
Then using the format
\be
A_{p}  \to  (v_1; v_2; v_3; v_4; v_5; v_6) 
\ee
we  may list their explicit values for $p=0,1,2,3,4,5$
\bea
A_0 &\to & 
\Big (1 ; 
{1\over 12} ; 
{1\over 720} ; 
-{1\over 720} ; 
{1\over 288} ; 
{ 1\over 120} \Big ) \ccr[3mm]
A_1 &\to & 
\Big (D-1; 
{D-7\over 12};
{D-16\over 720}; 
{91-D\over 720}; 
{D-13\over 288};
{ D-6\over 120} \Big ) \ccr[3mm]
A_2  &\to & 
\Big ({(D-1)(D-2)\over 2}; 
{D^2 -15 D + 38\over 24}; 
{(D-16)(D-17) \over 1440}; 
\ccr
&&   
-{D^2 -183 D + 1262\over 1440};
{D^2 -27 D + 146\over 576}; 
{D^2 -13 D + 32\over 240} \Big ) \ccr[3mm]
A_3  
&\to & 
\Big (
{1\over 6}(D-1)(D-2)(D-3);
{1\over 72}(D-3)(D^2 -21 D + 74);
\ccr
&&   
{1\over 4320}(D^3 -51 D^2 + 866 D -3246);
-{1\over 4320}(D^3 -276 D^2 + 4061 D -14046);
\ccr
&& {1\over 1728}(D-11)(D^2 -31 D + 138);
{1 \over 720} (D-3)(D^2 -18 D + 62)
\Big) \ccr[3mm]
A_4   
&\to &  
\Big ( 
{1\over 24}(D-1)(D-2)(D-3)(D-4);
\ccr
&&
{1\over 288}(D-3)(D-4)(D^2 -27 D + 122); 
\ccr 
&&    
{1\over 17280}(D^4 -70 D^3 + 1835 D^2 -14750D + 36024);
\ccr 
&& 
-{1\over 17280}(D^4 -370 D^3 + 8675 D^2 -64490D + 151224);
\ccr 
&&   {1\over 6912}(D^4 -58 D^3 + 1043 D^2 - 7058 D +15864); 
\ccr 
&& 
{1 \over 2880} (D-3)(D-4)(D-6)(D-17) 
\Big)   \ccr[3mm]
A_5   
&\to &  
\Big ( 
{1\over 120}(D-1)(D-2)(D-3)(D-4)(D-5);
\ccr 
&& 
{1\over 1440}(D-3)(D-4)(D-5)(D-7)(D-26);
\ccr 
&&    
{1\over 86400}(D-5)(D^4 -85 D^3 + 2810 D^2 -27500D + 81024);
\ccr 
&& 
-{1\over 86400}(D-5)(D^4 -460 D^3 + 13085 D^2 -117950 D + 334824); 
\ccr 
&&   {1\over 34560}(D-5)(D^4 -70 D^3 + 1535 D^2 - 12650 D + 34584); 
\ccr 
&& 
{1 \over 14400} (D-3)(D-4)(D-5)(D^2 -28 D + 152) 
\Big) \ .   
\eea 
They all satisfy the duality relations discussed in the text.


\subsection{Computation of the two-point function}
\label{sec:2point_app} 

The relevant contributions to the correlation function
 $\big \la e^{-S_{int}} \big \ra |_{\mbox{\tiny m.l.}}$ appearing in 
eq.~(\ref{master}) are the following 
\begin{alignat}{2}
\la e^{-S_{int}} \big \ra \Big|_{\mbox{\tiny m.l.}} 
&= -{1 \over 2\beta} \, \int_0^1 d\tau \la \,  c_{a\mu}c^a_\nu 
(\dot{y}^{\mu} \dot{y}^{\nu} + a^{\mu}a^{\nu} + b^{\mu} c^{\nu}) \, 
\ra \tag{A}
\\
&\phantom{=} 
-{1 \over \beta} \,\int_0^1 d\tau  \la \, \dot{y}^{\mu} 
\omega^{(2)}_{\mu ab} \bar{\psi}^a \psi^b \, \ra 
\tag{B} 
\\ 
&\phantom{=}
+{1 \over 2\beta} \,\int_0^1 d\tau  \la \, R^{(2)}_{abcd} 
\bar{\psi}^a \psi^b \bar{\psi}^c \psi^d \, \ra 
\tag{C}
\\ 
&\phantom{=}
+{1 \over 8\beta^2} \, 
\Big \la \, \left ( \int_0^1 d\tau \, 2 \, c_{(\mu \nu)} 
(\dot{y}^{\mu} \dot{y}^{\nu} + a^{\mu}a^{\nu} + b^{\mu} c^{\nu}) \right )^2 \,
\Big \ra 
\tag{D} 
\\
&\phantom{=}
+{1 \over 2\beta^2} \, \Big \la \, \left ( \int_0^1 d\tau \dot{y}^{\mu} 
\omega^{(1)}_{\mu ab} \bar{\psi}^a \psi^b  \right )^2 \, \Big \ra 
\tag{E} 
\\ 
&\phantom{=}
+{1 \over 2\beta^2} \, \Big \la \, \int_0^1 d\tau  \int_0^1 d\sigma \, 2 \,
 c_{(\mu \nu)} 
(\dot{y}^{\mu} \dot{y}^{\nu} + a^{\mu}a^{\nu} + b^{\mu} 
c^{\nu})(\tau) \dot{y}^{\mu} \omega^{(1)}_{\mu ab} \bar{\psi}^a \psi^b 
(\sigma) \, \Big \ra 
\tag{F} 
\\
&\phantom{=}
+{1 \over 8\beta^2} \, \Big \la \, \left ( \int_0^1 d\tau \, R^{(1)}_{abcd} 
\bar{\psi}^a \psi^b \bar{\psi}^c \psi^d 
\right )^2 \, \Big \ra 
\tag{G}
\\ 
&\phantom{=}
-{1 \over 4\beta^2} \,  \Big
\la \,  \int_0^1 d\tau \int_0^1 d\sigma \, 2 \, 
c_{(\mu\nu)}
(\dot{y}^{\mu} \dot{y}^{\nu} + a^{\mu}a^{\nu} + b^{\mu} c^{\nu})(\tau) 
R^{(1)}_{abcd} \bar{\psi}^a \psi^b \bar{\psi}^c \psi^d (\sigma) \, 
\Big \ra 
 \tag{H}
\\
&\phantom{=}
-{1 \over 2 \beta^2 } \, \Big \la \, \int_0^1 d\tau \int_0^1 d\sigma \, 
 \dot{y}^{\mu} \omega^{(1)}_{\mu ab} \bar{\psi}^a \psi^b (\tau)
 R^{(1)}_{cdef} \bar{\psi}^c \psi^d \bar{\psi}^e \psi^f (\sigma) \, \Big\ra 
\tag{I} \\
\label{correl}
\end{alignat}
where $R^{(i)}_{abcd}$ and $\omega^{(i)}_{\mu ab}$ denote the part of the
 Riemann tensor and spin connection with $i$ powers of the vielbein 
fluctuation $c_{a\mu}$. The first three contributions 
in eq.~(\ref{correl}) 
correspond to one-vertex graphs, while the remaining six 
contributions  correspond to two-vertex graphs (see figure 3). 
The calculation is a straightforward generalization of 
the analogous ones performed in \cite{Bastianelli:2002fv} and 
\cite{Bastianelli:2002qw} for the scalar and the fermion. 
It is convenient to use the following string inspired propagators
\bea
\la y^\mu (\tau) y^\nu(\sigma)\ra  &=& 
- \beta  \delta^{\mu\nu} 
\Delta(\tau-\sigma)
\nonumber\\
\la a^\mu(\tau) a^\nu(\sigma)\ra &=& 
  \beta  \delta^{\mu\nu} \Delta_{gh} (\tau- \sigma) 
\nonumber\\
\la b^\mu(\tau) c^\nu (\sigma)\ra  &=&  -2\beta \delta^{\mu\nu} 
\Delta_{gh}(\tau-\sigma)
\nonumber\\
\la \psi^a(\tau) \bar \psi_b(\sigma)\ra 
&=& \beta \delta^a_b \Delta_{AF}(\tau-\sigma)
\eea
where $\Delta$, $\Delta_{gh}$ and $\Delta_{AF}$
are given by
\bea
&& \Delta (x) = -\sum_{n\neq0}
 {1 \over {4 \pi^2 n^2}} e^{2 \pi i  n x }
={1\over 2} |x| -
{1\over 2} x^2 -{1 \over 12}
\nonumber\\
&& \Delta_{gh}(x) = \sum_{n=-\infty}^{\infty}
e^{2 \pi i  n x} = \delta(x) 
\nonumber\\
&& \Delta_{AF}(x)=
\sum_{r\in Z+{1\over 2}} {-i\over 2 \pi  r + \phi }\,  e^{2  \pi i r x} 
= {e^{-i\phi x} \over 2 \cos {\phi\over 2}}
\Big [ e^{i{\phi\over 2}}\theta(x)  - e^{-i{\phi\over 2}}\theta(-x)\Big ] \ ,
\eea
and dimensional regularization as described in paper I.

To organize the 
computation, it is convenient to introduce the following basis of tensors 
\bea
R_1^{\mu\nu\alpha\beta} \!\! &=& \!
\delta^{\mu\nu}\delta^{\alpha\beta}
\nonumber \\
R_2^{\mu\nu\alpha\beta} \!\! &=& \!
\delta^{\mu\alpha}\delta^{\nu\beta}+ \delta^{\mu\beta}\delta^{\nu\alpha} 
\nonumber \\
\bar R_2^{\mu\nu\alpha\beta} \!\! &=& \!
\delta^{\mu\alpha}\delta^{\nu\beta}- \delta^{\mu\beta}\delta^{\nu\alpha} 
\nonumber \\
R_3^{\mu\nu\alpha\beta} \!\! &=& \!
{1\over k^2} \, (\delta^{\mu\alpha} k^\nu k^\beta +
\delta^{\nu\alpha} k^\mu k^\beta +
\delta^{\mu\beta} k^\nu k^\alpha +
\delta^{\nu\beta} k^\mu k^\alpha ) \nonumber\\
R_4^{\mu\nu\alpha\beta} \!\! &=& \!
 {1\over k^2} \, (\delta^{\mu\nu} k^\alpha k^\beta
+\delta^{\alpha\beta} k^\mu k^\nu)
\nonumber \\
R_5^{\mu\nu\alpha\beta} \!\! &=& \!
 {1\over k^4}\, k^\mu k^\nu k^\alpha k^\beta~.
\label{r-tensors_app}
\eea
It will be also useful to define the following transverse combinations
\bea
S_1^{\mu\nu\alpha\beta} \!\! &=& \!\! \left 
( R_1 -  R_4 + R_5 \right )^{\mu\nu\alpha\beta} \nonumber \\
S_2^{\mu\nu\alpha\beta} \!\! &=& \!\! \left 
( R_2 -  R_3 + 2 R_5 \right )^{\mu\nu\alpha\beta} .
\label{s-tensors_app}
\eea
In the following, we will suppress the explicit polarization tensors by use 
of the notation $R_i = \epsilon^{(1)}_{\mu\nu} 
\epsilon^{(2)}_{\alpha\beta} R_i^{\mu\nu\alpha\beta}$.
 Explicitly, we get the following results:
\begin{itemize}
\item (A) was already computed for the $N=1$ sigma
 model \cite{Bastianelli:2002qw}, and is equal to
\bea 
(A) = \frac{1}{2} (R_2 + \bar{R}_2)  \ .
\eea
\item (B) vanishes being proportional to $\omega_{\mu ab} \delta^{ab}$.
\item (C) is given by
\bea 
(C) = {1 \over 2\beta} \,\int_0^1 d\tau  \la \,
 R^{(2)}_{abcd} \bar{\psi}^a \psi^b \bar{\psi}^c \psi^d \, \ra 
= {\beta \over 2} \Delta_{AF}^2(0) \int_0^1 d\tau  \la \, R^{(2)} \, \ra \ .
\eea
The Ricci-scalar correlation function has already been computed 
for the scalar field in the $N=0$ sigma model \cite{Bastianelli:2002fv}, 
using the metric basis. The result in the vielbein basis is four times 
that result
\bea 
\la \, R^{(2)} \, \ra = -k^2 (2R_1 +R_2 -R_3) \ .
\eea
So we obtain
\bea
(C) = { \beta k^2 \over 8} \,  \mbox{tan}^2 {\phi \over 2}  \,
 (2R_1 +R_2 -R_3)  \ .
\eea
\item (D) has also been computed already for the scalar
 \cite{Bastianelli:2002fv} and the fermion \cite{Bastianelli:2002qw} 
running in the loop. The result in the vielbein basis is just four times
 the result in the metric basis, which was presented in detail 
in \cite{Bastianelli:2002fv}
\bea
(D) &=& {1 \over 8\beta^2} \, \Big
\la \, \left ( \int_0^1 d\tau \, 2 \,
 c_{(\mu \nu)} 
(\dot{y}^{\mu} \dot{y}^{\nu} + a^{\mu}a^{\nu} + b^{\mu} c^{\nu}) \right )^2 \,
 \Big \ra \nonumber \\
&=& A_0 (S_1+S_2) +{\beta k^2 \over 8} S_2 + (R_3+R_4-2R_2-3R_5) 
\eea
where 
\bea
A_0 = \int_0^1 d\tau e^{-{\beta \over 2}k^2 (\tau-\tau^2)} \  .
\eea
\item The computation of (E) is very similar to the computation of the
analogous term in the $N=1$ case. It needs dimensional regularization (DR) 
and the final result turns out to be
\bea
(E) = {1 \over 2\beta^2} \, \Big 
\la \, \left ( \int_0^1 d\tau \dot{y}^{\mu} 
\omega^{(1)}_{\mu ab} \bar{\psi}^a \psi^b  \right )^2 \, \Big   \ra = 
{\beta k^2 \over 4 \mbox{cos}^2 {\phi \over 2}} S_2 (A_0-1) \ .
\eea 

\item (F) vanishes for the same reason as (B).
\item (G) can be computed by first contracting all the fermions
\bea
(G) &=& {1 \over 8\beta^2} \, \Big \la \, \left ( \int_0^1 d\tau \,
 R^{(1)}_{abcd} \bar{\psi}^a \psi^b \bar{\psi}^c \psi^d 
\right )^2 \, \Big \ra \nonumber \\
&=& {\beta^2 \over 8} \int_0^1 d\tau \int_0^1 d\sigma \, 
 \bigg{(} \Delta_{AF}^4 (0) \la \, R^{(1)}(\tau)R^{(1)}(\sigma) \, \ra
\nonumber \\
&&-4 \Delta_{AF}^2 (0) \Delta_{AF}(\tau-\sigma) \Delta_{AF}(\sigma-\tau) \la \,
 R^{(1)}_{ab}(\tau)R^{(1)}_{ab}(\sigma) \, \ra \nonumber \\
&& + \Delta_{AF}^2(\tau-\sigma) \Delta_{AF}^2(\sigma-\tau) \la \,
 R^{(1)}_{abcd}(\tau)R^{(1)}_{abcd}(\sigma) \, \ra \bigg{)} \ .
\eea
The remaining correlation functions yield
\bea
\la \, R^{(1)}(\tau)R^{(1)}(\sigma) \, \ra &=& 8 k^4 S_1
 e^{-\beta k^2 \Delta_0(\tau-\sigma)} \nonumber
\\
\la \, R^{(1)}_{ab}(\tau)R^{(1)}_{ab}(\sigma) \, \ra &=& k^4 (2S_1 + S_2) 
e^{-\beta k^2 \Delta_0(\tau-\sigma)} \nonumber\\
\la \, R^{(1)}_{abcd}(\tau)R^{(1)}_{abcd}(\sigma) \, 
\ra &=& 4 k^4 S_2 e^{-\beta k^2 \Delta_0(\tau-\sigma)}  \ ,
\eea
where $\Delta_0(\tau-\sigma)=\Delta(\tau-\sigma) - \Delta(0)$. 
The resulting worldline integrals do not need DR, and the final result is
\bea
(G) &=& {\beta^2 k^4 \over 16} \left ( \mbox{tan}^4 {\phi \over 2} \,
 S_1 - {\mbox{tan}^2 {\phi \over 2} \over 2 \mbox{cos}^2 {\phi \over 2}}
 \, (2S_1+S_2) + {1 \over 2 \mbox{cos}^4 {\phi \over 2}} \, S_2 \right )
 A_0 \nonumber \\ 
&=& {\beta^2 k^4 \over 16} \left (S_1 + {1 \over 2 \mbox{cos}^2 
{\phi \over 2}} \, (S_2-2 S_1) \right ) A_0 \ .
\eea
Note that the terms proportional to $\mbox{cos}^{-4} {\phi \over 2}$
 cancel out identically. This is going to be important for the expected 
duality of the final result.
\item After contracting the four fermions, (H) becomes
\bea
(H) =  -{1 \over 4} \,    \int_0^1 d\tau \int_0^1 d\sigma \la \, \, 2 \,
 c_{(\mu\nu)}
(\dot{y}^{\mu} \dot{y}^{\nu} + a^{\mu}a^{\nu} + b^{\mu} c^{\nu})(\tau) 
R^{(1)} \, \ra \Delta_{AF}^2(0)  \ . \qquad 
\eea
The remaining correlation function was already computed for the scalar in
 the $N=0$ model, the result in the vielbein basis is 4 times that result,
 and we get
\bea
(H) = -{\beta k^2 \over 2} \mbox{tan}^2 {\phi \over 2} 
\left ( A_0 S_1 -R_5 + {1 \over 2} R_4 \right ) \ .
\eea
\item Finally, (I) vanishes being proportional to $\omega_{\mu ab} R_{ab}$.
\end{itemize}
Summing up all the contributions in eq.~(\ref{correl}), we get the expression
in eq. (\ref{final}).

\vfill\eject


\end{document}